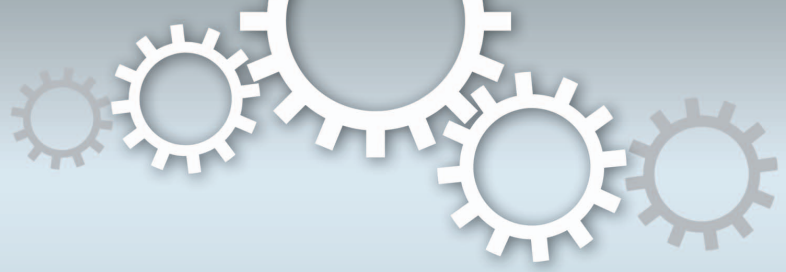

# Correlated Diskoid-like Electronic States

Artem Baskin[1], Hossein R. Sadeghpour[2] & Petr Král[1,3]

[1]Department of Chemistry, University of Illinois at Chicago, Chicago, IL 60607, USA, [2]ITAMP, Harvard-Smithsonian Center for Astrophysics, Cambridge, Massachusetts 02138, [3]Department of Physics, University of Illinois at Chicago, Chicago, IL 60607, USA.

We study highly excited diskoid-like electronic states formed in the vicinity of charged and strongly polarizable diskotic nanostructures, such as circular graphene flakes. First, we study the nature of such extended states in a simple two-electron model. The two electrons are attached to a point-like nucleus with a charge 2+, where the material electron is forced to move within a 2D disk area centered at the nucleus, while the extended electron is free to move in 3D. Pronounced and complex correlations are revealed in the diskoid-like states. We also develop semiclassical one-electron models of such diskotic systems and explain how the one-electron and many-electron solutions are related.

H ighly extended Rydberg-like electronic states in atoms and molecules[1,2] bear semi-classical concepts developed in the Bohr model. Molecular Rydberg states can have extremely large permanent electric dipole moments[1] highly sensitive to their environment, which is useful in probing physical and chemical processes at the nanoscale[3,4]. These extended electronic states can provide unique information about light-matter interactions[5,6] and particle entanglement[7,8], with implications in quantum information and metrology sciences.

Similar extended electronic states were predicted to form[9–11] and observed[12,13] above the surfaces of metallic carbon nanotubes (CNT). In these tubular image state (TIS), the electron is attracted to its hole image formed on a highly (longitudinally) polarizable CNT surface. In principle, analogous diskoid-like electronic states (DES) could be formed above highly polarizable diskotic nanostructures, such as graphene flakes or metallic nanodisks. However, in contrast to TISs, here the image hole is induced in a transversely polarizable disk, which means that it should be highly correlated with the orbiting electron. In this work, we introduce such highly excited diskoid-like electronic states, find their solutions in a two-electron and one-electron (mean field) approximations, and discuss their properties, in particular their correlations.

## Results: Two-electron Diskoid-like States

A fully quantum mechanical description of correlated electronic states formed above nanoscale metallic surfaces requires a proper accounting of many-body effects[14]. However, finding accurate solutions of the Schrödinger equation for few electrons moving in arbitrary confining potentials remains a highly challenging task. The problem of a pair of interacting electrons confined in a 3D cavity was solved exactly only recently for certain types of cavity geometries[15–17]. Therefore, the development of simplified models where we could track, in a semi-analytical fully explicit way, the emergence of electron-electron correlations in few-electron systems may have a potential impact on several scientific areas.

We start with our discussion of DESs in one of the simplest possible approximation, where correlations between the external electron and material electrons can be clearly observed. We introduce a simple two-electron DES model, schematically shown in Fig. 1. The two electrons move around a localized nucleus with a charge of $Z = +2$, where the "material" (polarization) electron (ME) is confined in a 2D disk of the radius $a$ ($z = 0$, $\rho \leq a$), whereas the "external" electron (EE) is free to move in a 3D space.

**Theoretical model.** The "hybrid" He atom in Fig. 1 can be described by the Hamiltonian (a.u.)

$$
\begin{aligned}
H = &-\frac{1}{2}\left(\frac{\partial^2}{\partial \rho_1^2} + \frac{1}{\rho_1}\frac{\partial}{\partial \rho_1} + \frac{1}{\rho_1^2}\frac{\partial^2}{\partial \vartheta^2}\right) - \frac{Z}{\rho_1} \\
&-\frac{1}{2}\left(\frac{\partial^2}{\partial \rho_2^2} + \frac{1}{\rho_2}\frac{\partial}{\partial \rho_2} + \frac{1}{\rho_2^2}\frac{\partial^2}{\partial \vartheta^2} + \frac{\partial^2}{\partial z_2^2}\right) - \frac{Z}{\sqrt{\rho_2^2 + z_2^2}} \\
&+ \frac{L_z^2}{2(\rho_1^2 + \rho_2^2)} + \frac{1}{\rho_{12}},
\end{aligned}
\tag{1}
$$

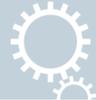

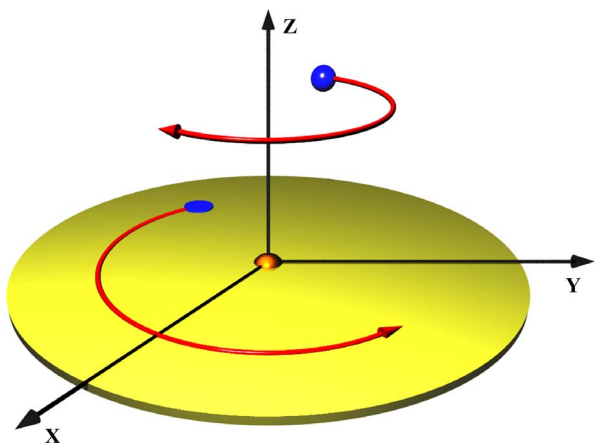

**Figure 1 | Model (He-type) system with diskoid-like states, where one electron is confined within the 2D disk area in the $z = 0$ plane and the second electron is orbiting around the disk in 3D.**

written in the relative and position-weighted angular coordinates[18], $\delta = \phi_1 - \phi_2$, $\quad \phi = \frac{\rho_1^2 \phi_1 + \rho_2^2 \phi_2}{\rho_1^2 + \rho_2^2}$, related to the two-electron coordinates, $(\rho_1, \phi_1)$, $(\rho_2, \phi_2, z_2)$, where $\rho_{12} \equiv \sqrt{\rho_1^2 + \rho_2^2 + z_2^2 - 2\rho_1\rho_2 \cos(\delta)}$ and $Z$ is the nucleus charge. The first and second lines in Eqn. 1 are the one-electron $h_1(\rho_1, \delta)$ (ME) and $h_2(\rho_2, z_2, \delta)$ (EE) Hamiltonians in effective cylindrical coordinates, respectively. The third line represents the Coulomb coupling of the two electrons and their centrifugal term, with $L_z^2 = -\frac{\partial^2}{\partial \phi^2}$.

We find the correlated two-electron eigenstates of $H$ in Eqn. 1 by a configuration interaction (CI) method, using a variational technique with a one-electron basis set wavefunctions, which solve $h_1(\rho_1, \delta)$ and $h_2(\rho_2, z_2, \delta)$. The cylindrical symmetry of the Hamiltonian, associated with the point group $D_{\infty h}$, prevent us from using the Hylleraas basis set[19,20] (He atom problems). Moreover, the mixed dimensionality of our problem necessitates to consider ME and EE as distinguishable Coulombically coupled particles. Then, the CI eigenfunctions can be written in a separable form, $\Psi(\vec{r}_1, \vec{r}_2) = \Psi_\Lambda(\rho_1, \rho_2, z_2, \delta)\frac{e^{\pm i\Lambda\phi}}{\sqrt{2\pi}}$, where $\Lambda$ is an eigenvalue of $L_z$. We will focus on $\Sigma$ states with $\Lambda = 0$ and expand $\Psi_0$ in the eigenstates of $h_{1,2}$.

As the Coulomb interactions of the ME and EE with nucleus do not depend on $\delta$, this angle can be factored out in both ($h_{1,2}$) sub-systems, as $\psi(\delta) = \frac{1}{\sqrt{2\pi}}e^{\pm il\delta}$. In the solution of the 2D H-atom[21], described by $h_1(\rho_1, \delta)$, the radial part of the eigenfunctions can be expressed in terms of confluent hypergeometric functions,

$$R_{1nl}(\rho_1) = {}_1F_1(-n + |l| + 1, 2|l| + 1, \beta_n\rho_1)$$
$$\times A_{nl}(\beta_n\rho_1)^{|l|}e^{-\beta_n\rho_1/2}, \quad E < 0,$$
$$R_{1kl}(\rho_1) = {}_1F_1(i/k + |l| + 1/2, 2|l| + 1, i2k\beta\rho_1) \quad (2)$$
$$\times C_{kl}(2k\beta\rho_1)^{|l|}e^{-ik\beta\rho_1}, \quad E > 0,$$

where $\beta_n = \frac{2Z}{n - 1/2}\frac{1}{a_0}$, $\beta = \frac{Z}{a_0}$, $k = \left(\frac{2E}{Z^2}\frac{a_0}{e^2}\right)^{1/2}$, and $a_0$ is the Bohr radius. The confinement of the ME in the disk area can be resolved by applying a Dirichlet boundary condition, $R_{1nl}|_{\rho_1 = a} = 0$[22,23], which selects and restricts the wavefunctions in Eqn. 2.

Then, the allowed energies of ME can be found from ${}_1F_1(-\zeta + |l| + 1, 2|l| + 1, a) = 0$. For a given value of $|l|$, the first root corresponds to the ME energy of the lowest ($n = |l| + 1$) state and the successive roots are its excited states. The confinement also removes the $l$-degeneracy, leaving only the twofold degeneracy with respect to the sign of $l$, so the energy spectrum can no longer be expressed as in the free 2D H atom, $E_n = -\frac{Z^2}{2(n - 1/2)^2}$. When the disk radius $a$ is decreased, the ME energy increases, passes through zero, and rapidly rises to large values. In Table I, we summarize the energies for 2D H-atom confined in the disk of radius $a = 1$ nm. The 3D H-atom Hamiltonian, $h_2(\rho_2, z_2, \delta)$, gives the solution for EE in the cylindrical coordinates, $R_{2nl}(\rho_2, z_2)$[24].

Using these one-electron solutions, the combined CI wavefunctions can be expanded as $\Psi_\Lambda^N(\rho_1, \rho_2, z_2, \delta) = \sum_{nkl} C_{nkl}^N R_{1nl}(\rho_1) \times R_{2kl}(\rho_2, z_2)\psi_l(\delta)$, where we omit the permutational spin symmetry, due to the distinguishability of ME and EE. However, these wavefunctions should have spatial symmetries belonging to the $D_{\infty h}$ group, i.e., the $i$ inversion in the force center and the $\sigma_v$ reflection in the plane containing $z$-axis.

These symmetry conditions can be satisfied by selecting certain $\Psi_\Lambda^N$-wavefunction components. We need to take into account that $R_1(\rho_1)$ is totally symmetric, $R_2(\rho_2, z_2)$ is gerade/ungerade with respect to $i$, and linear combinations of $\psi_l(\delta)$ can be either symmetric or antisymmetric with respect to $\sigma$. Then, considering $\Sigma_{g/u}^\pm$ ($\Lambda = 0$), the allowed symmetry-adapted wavefunction products are $A_{g/u}^{N, \pm} R_{1nl}(\rho_1) R_{2kl}^{g/u}(\rho_2, z_2)\psi_l^\pm(\delta)$. Here, $\psi_l^+(\delta) = \frac{1}{\sqrt{\pi(1 + \delta_{l0})}}\cos(l\delta)$ for $l = 0, 1, 2, \ldots$ and $\psi_l^-(\delta) = \frac{1}{\sqrt{\pi}}\sin(l\delta)$ for $l = 1, 2, 3 \ldots$ and $A_{g/u}^\pm$ are the normalization coefficients. For simplicity, we focus on the $\Sigma_g^+$ state and use a limited basis set which consists of all the possible combinations of eigenfunctions of $h_{1,2}$, with $l = 0, 1, 2, \ldots, 9$ and $n = l + 1$,

**Table I | The energy eigenvalues (a.u.) for orbitals of the 2D hydrogen atom ($Z = 2$) confined in the disk of radius $a = 1$ nm. Energy states with $E < 0$ and $E > 0$ are highlighted with blue and red colors, respectively**

| $l = 0$ | $l = 1$ | $l = 2$ | $l = 3$ |
|---|---|---|---|
| $E_{10} = -8.005643$ | $E_{21} = -0.889515$ | $E_{32} = -0.163121$ | $E_{43} = -0.01118$ |
| $E_{20} = -0.889515$ | $E_{31} = -0.320225$ | $E_{42} = -0.082219$ | $E_{53} = 0.100311$ |
| $E_{30} = -0.320225$ | $E_{41} = -0.162923$ | $E_{52} = 0.007547$ | $E_{63} = 0.242860$ |
| $E_{40} = -0.162837$ | $E_{51} = -0.080790$ | $E_{62} = 0.133351$ | $E_{73} = 0.415372$ |

| $l = 4$ | $l = 5$ | $l = 6$ | $l = 7$ |
|---|---|---|---|
| $E_{54} = 0.189837$ | $E_{65} = 0.448944$ | $E_{76} = 0.766247$ | $E_{87} = 1.142297$ |
| $E_{64} = 0.347205$ | $E_{75} = 0.650439$ | $E_{86} = 1.011136$ | $E_{97} = 1.430107$ |
| $E_{74} = 0.534240$ | $E_{85} = 0.880975$ | $E_{96} = 1.284670$ | $E_{107} = 1.746312$ |
| $E_{84} = 0.750451$ | $E_{95} = 1.140261$ | $E_{106} = 1.586679$ | $E_{117} = 2.090793$ |

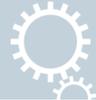

…, $l + 4$ eigenvalues. Overall, the basis set contains 160 basis functions.

First, we discuss the *angular* distributions in the two-electron wavefunctions (correlations) for ME and EE, caused by the electron-electron coupling. Figure 2 shows the spatial distribution of ME, $\rho_1\rho_2|\Psi_{\Sigma_g}(\vec{r}_1,\vec{r}_2)|^2$, in different excited $\Sigma_g$ states, when EE is positioned at a distance of $\rho_2 = 0.2\,a$ and $\rho_2 = 2.5\,a$ from the center and $\phi_2 = 0$ (the $x$-axis in the disk plane). We can see that all the considered states ($N = 5, 29, 48, 120$) have a significant angular asymmetry of the ME distribution. Moreover, the radial nodal pattern seems to be also altered in the moderately excited (b, c) states. In the higher excited state (d), one would expect that an opposite charge (hole) is accumulated close to the external electron in the disk. However, since a single ME can not represent well a metal, this hole does not form.

Figure 3 (top) shows the evolution of the ME distribution in $N = 29$ for different radial positions of EE, $\rho_2 = 0.1 - 0.8\,a$ ($\Delta\rho_1 = 0.1a$), localized in the disk plane ($z_2 = 0, \phi_2 = 0$). As $\rho_2$ grows oscillations of the ME density can be observed (correlations with EE). Figure 3 (middle) shows the same for the $N = 120$ state, but $\rho_2 = a - 3.8\,a$ and $\Delta\rho_2 = 0.4\,a$. We can observe very interesting oscillations in the ME density even when EE is far away from the disk. Figure 3 (bottom) presents the EE distribution for $N = 72$ outside the disk area (gray), when the ME position is changing like that of EE in Fig. 3 (top). This time, we observe oscillations of the EE density localized near the disk edge, where most of the EE population is largely present. However, the angular asymmetry is not seen farther from the disk.

We also study the *radial* distributions in the same two-electron wavefunctions for ME and EE. Figure 4 (top) compares the one-electron density matrix for ME, $\rho_1|\bar{\Phi}_N(\rho_1)|^2 = \rho_1\int\rho_2$ $\times|\Phi_N(\rho_1,\rho_2,z_2,\delta)|^2 d\rho_2\,dz_2\,d\phi_2$, with the one-electron distribution, $\rho_1|R_{1l_1n_1}(\rho_1)|^2$, where $l_1, n_1$ correspond to the basis function which has the largest variational coefficient in $\Phi_N(\rho_1, \rho_2, z_2, \delta)$. In the middle excited states ($N = 29, 48$) the correlations significantly perturb the ME distribution. In these states, $\rho_1|\bar{\Phi}_N(\rho_1)|^2 \neq 0$ is largely spread over the whole disk, where even the nodal pattern is suppressed, as noticed in Fig. 2. However, in contrast to angular correlations (Fig. 2), the radial correlations are suppressed in the low ($N = 5$) and highly ($N = 120$) excited states.

In Fig. 4 (bottom), we perform the same analysis for EE. In particular, we compare $\rho_2|\Phi_N(\rho_2,z_2)|^2 = \rho_2\int\rho_1|\Phi_N(\rho_1,\rho_2,z_2,\delta)|^2$ $\times d\rho_1\,d\phi_1$ with $\rho_2|R_{2l_2n_2}(\rho_2,z_2)|^2$, where we set $z_2 = 0$ (EE in disk plane) and pick $l_2, n_2$ for the largest variational coefficient in $\Phi_N(\rho_1, \rho_2, z_2, \delta)$. The radial distributions of EE are affected for all the chosen states, but for $N = 29, 48$ the correlations are substantially larger.

In order to analyze the electron correlations in more detail, we evaluate how the eigenstates $\Phi_N$ are spread over the used basis wavefunctions, and compare these results with the average ME-EE distance $\langle r_{12}\rangle$ in these states. In Fig. 5 (top), we present the standard deviation of the variational coefficients for $N = 1$–160 states, normalized as $\sum_i^M |c_i^N|^2 = 1$. We evaluate the degree of delocalization from $\Delta_{loc} = 1 - \sqrt{1 - \left(\sum_i^M |c_i^N|^2\right)^2 / M}$. For low and highly excited states the eigenfunctions are localized on a few one-electron wavefunctions, whereas the eigenfunctions in the middle part of the energy spectrum are more evenly distributed over the basis set. For each $\Phi_N$, we also calculate the average number of states $\langle n\rangle/M$ whose variational coefficients exceed a threshold $\epsilon$. In Fig. 5 (top, inset) the distribution shows how many basis wavefunctions contribute to the $N$ state at a given threshold $\epsilon$. For the chosen thresholds $\epsilon = 0.02$, 0.05, the eigenfunctions have practically the same profiles of delocalization as when obtained through standard deviation of the variational coefficients.

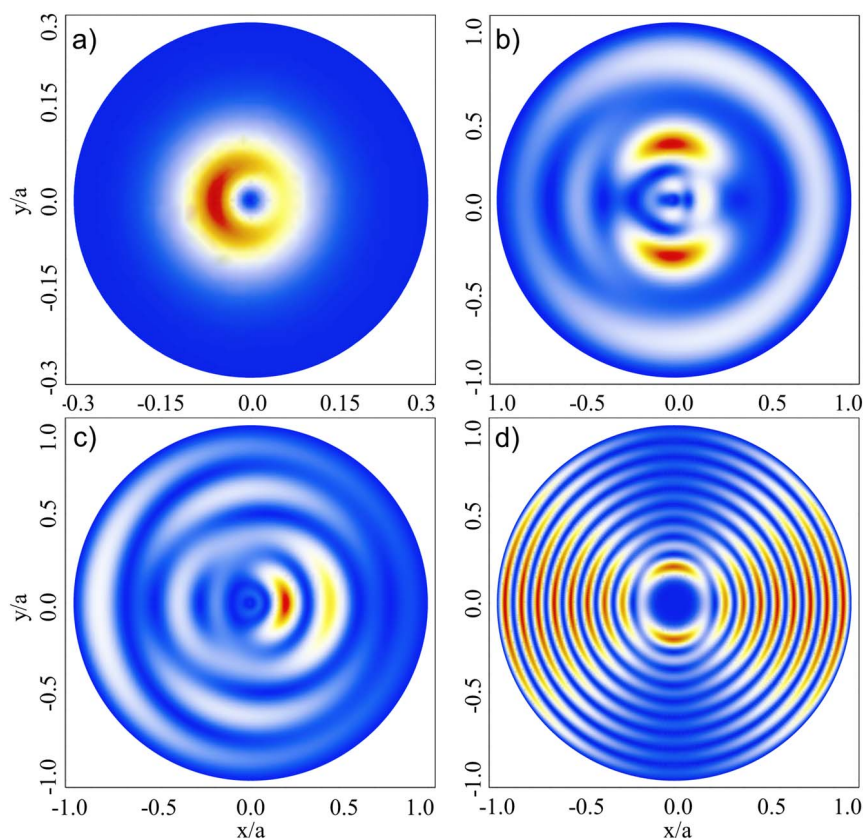

**Figure 2 | ME density $\rho_1\rho_2|\Psi_{\Sigma_g}(\vec{r}_1,\vec{r}_2)|^2$ ($\vec{r}_2$ fixed) when EE sits on the disk plane at a distance of $\rho_2 = 0.2\,a$ (cases a, b, c) and $\rho_2 = 2.5\,a$ (case d) from the center with $\phi_2 = 0$ for a) $N = 5$, b) $N = 29$, c) $N = 48$, d) $N = 120$. The coordinates are given in units of a disk radius, $a = 1$ nm.**

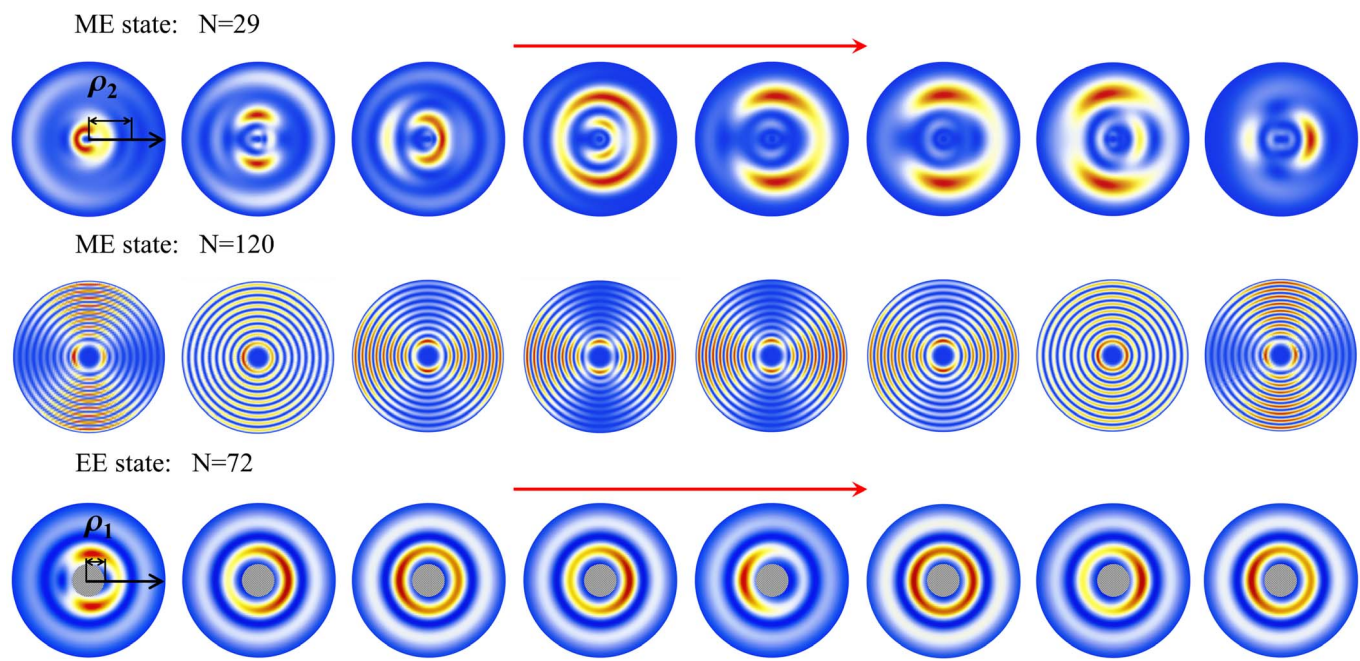

**Figure 3** | (top) Evolution of the ME density $\rho_1\rho_2|\Psi_{\Sigma_g}(\vec{r}_1,\vec{r}_2)|^2$ ($\vec{r}_2$ fixed) for EE positioned in the disk plane ($z_2 = 0$, $\phi_2 = 0$) at different distances $\rho_2$ from the disk center for the $N = 29$ state. The snapshots correspond to $\rho_2 = 0.1$ a–0.8 a values separated by $\Delta\rho_2 = 0.1$ a. The area $\rho_1 \leq a$ is shown. (middle) The same as in (top) for the $N = 120$ state. The snapshots correspond to $\rho_2 = 1$ a–3.8 a values separated by $\Delta\rho_2 = 0.4$ a. The area $\rho_1 \leq a$ is shown. (bottom) Evolution of the EE density $\rho_1\rho_2|\Psi_{\Sigma_g}(\vec{r}_1,\vec{r}_2)|^2$ ($\vec{r}_1$ fixed) in the disk plane ($z_2 = 0$) for different distances $\rho_1$ of ME ($\phi_1 = 0$) from the center of the $N = 72$ state. The snapshots correspond to $\rho_1 = 0.1$ a–0.8 a values separated by $\Delta\rho_1 = 0.1$a. The area $\rho_2 \leq 4a$ is shown. The disk is gray.

In Fig. 5 (bottom), we show that the $\langle r_{12} \rangle$ dependence on $N$ is almost linear; the quasi-periodic oscillations originate from the specific choice of the basis set. We found that $\langle r_{12} \rangle$ has local maxima in the $N$ states which are localized on a particular basis wavefunction and it has local minima in the $N$ states which are more delocalized over the basis set.

These observations agree with the previous results, since in the $N = 29$, 48 states, where $\langle r_{12} \rangle$ has a minimum, ME has a highly correlated radial distribution, seen in Fig. 2 b–c and 4. The suppressed radial correlations in the low excited stated can be related to the mixed dimensionality of this problem, where ME coupling to the nucleus can not be well disturbed by the electron repulsion with EE. In moderately excited states, $\langle r_{12} \rangle \sim 1$–2 nm and the mixed dimensionality becomes less relevant. In highly excited states, $\langle r_{12} \rangle \simeq 3$–5 nm, so the radial correlations largely disappear.

These electron-electron correlations in DES could potentially be observed experimentally using spectroscopic techniques. One could measure the energy shifts in optical transitions (compared to theoretical spectra of non-correlated system) or reductions of electron kinetic energies by the Coulomb repulsion[25]. Because of the highly correlated nature of DESs, the highly excited electron orbiting such discoid structures might be prone to decay (angular momentum exchange with material electrons). This electron-electron damping mechanism was much less significant in TISs, where the electron coupling to nanotube vibrations was the dominant damping mechanism[10].

**Mean-field solutions.** In principle, we can describe at a semiclassical (one-electron) level the motion of highly excited electrons moving in a mean-field potential formed above diskoid nanostructures. Following Fig. 1, we assume that EE moves around a charged and polarizable (ME) disk ("Hartree hole" of an image charge). In principle, these semi-classical problems can be studied by asymptotic expansion methods[20,21,26–30] or perturbation methods, which can be used to obtain one-electron solutions of this mean-field problem[26,28].

We illustrate two ways of finding effective mean-field potentials, $U_N(\rho_2, z_2)$, which could describe the interaction of EE with the nucleus and ME (affected by EE). In the first case, we evaluate the average potential that EE experiences, provided that its angular momentum quantum number is $l_2$ and ME is in the $|l_1, n_1\rangle$ state, $U^I_{l_1,l_2,n_1}(\rho_2, z_2) = \frac{l_2^2 - 1/4}{\rho_2^2} - \frac{2}{\sqrt{\rho_2^2 + z_2^2}} + \frac{1}{2\pi}\int R_{1l_1n_1}^2(\rho_1)\frac{\rho_1}{\rho_{12}}\,d\rho_1\,d\delta$. This potential does not include the polarization of ME, but the nucleus screening is accounted for. Alternatively, we can obtain the mean-field potential, $U^{II}_N(\rho_2, z_2) = \frac{l_2^2 - 1/4}{\rho_2^2} - \frac{2}{\sqrt{\rho_2^2 + z_2^2}} + \int |\bar{\Phi}_N(\rho_1)|^2 \frac{\rho_1}{\rho_{12}}\,d\rho_1\,d\delta$, where $|\bar{\Phi}_N(\rho_1)|^2 = \int |\Phi_N(\rho_1, \rho_2, z_2, \delta)|^2 \rho_2\,d\rho_2\,dz_2\,d\phi_2$. This is an effective potential between EE, the force center and ME with a distribution affected by EE.

In Fig. 6, we present the $U^I$ and $U^{II}_N$ potentials. The $U^I_{l,n}(\rho_2)$ potential is calculated for different $l_1 = l_2$ and $n_1 = l_1 + 1$. It develops a well detached local minimum for $l_2 \geq 5$ at $\rho_2 > 1.5$ a. Small ripples in the potential form at $\rho_2 < a$ due to an inhomogeneous charge distribution on the disk. The ripples depend slightly on $n_1$, but vanish at $l_2 > 6$. The quantum numbers $l_1, l_2, n_1$ are extracted from the CI coefficients of the "exact wavefunction". However, the correspondence between $|\Lambda = 0, N\rangle$ and $|l_1, l_2, n_1\rangle$ is approximate. Comparison of the potentials shows that polarization is important for $l_2 = 2$, 3. This is compatible with our exact two-electron solution where correlation effects are negligible for low excited states rates in $l_2 = 0$, 1 (Fig. 4 a). Both mean-field and classical (Fig. 8 (top)) potentials develop well detached minima for $l_2 \geq 5$ at $\rho_2 > 1.5$ a and $l \geq 8$ at $\rho > 3.5$ a, respectively. For highly excited states, both the disk-electron distance and the average electron-electron distance are large, so the correlations are no longer seen.

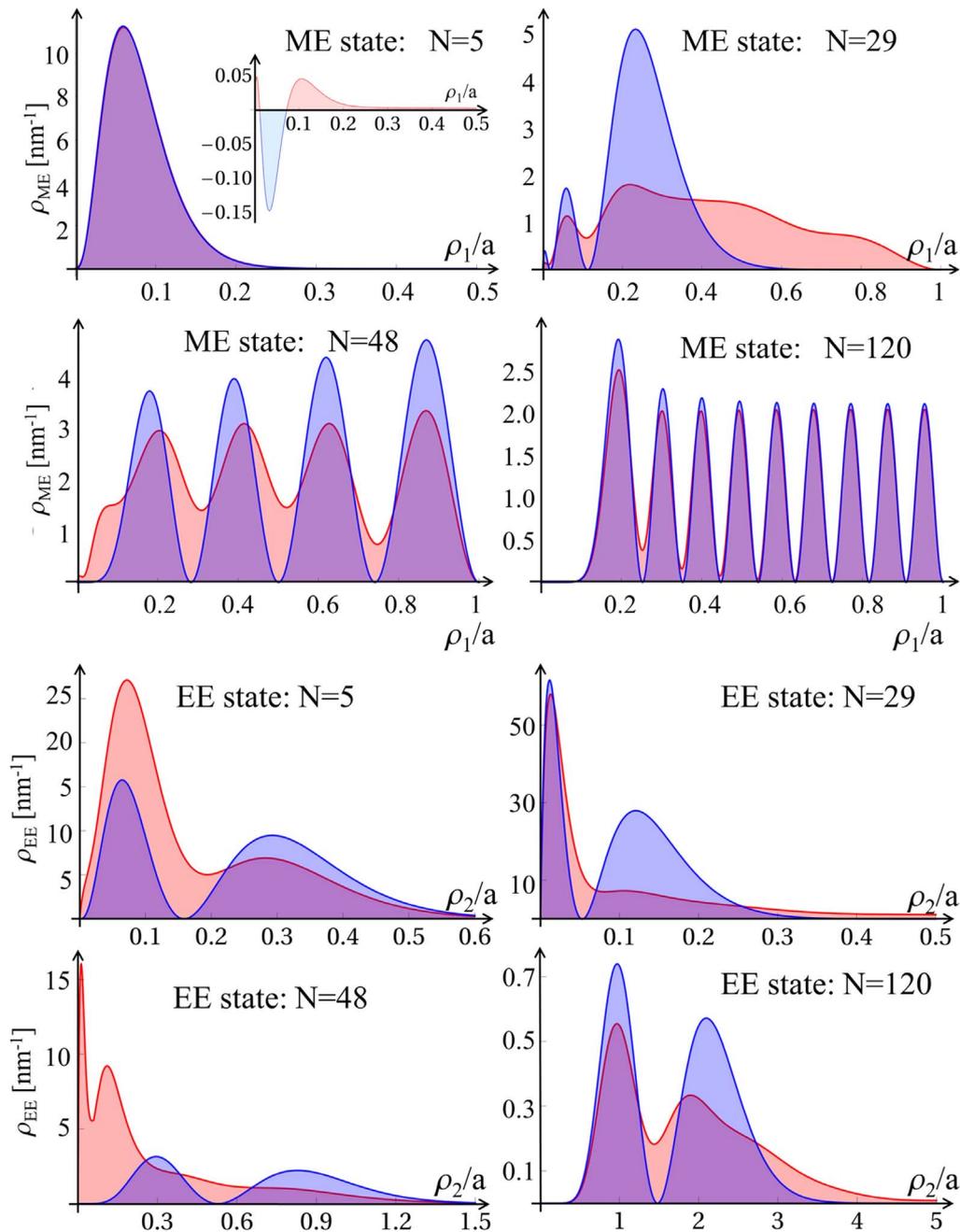

**Figure 4** | (top) The radial distributions $\rho_{ME}$ of ME, $\rho_1 |R_{1l_1n_1}(\rho_1)|^2$ and $\rho_1 |\tilde{\Phi}_N(\rho_1)|^2$, obtained from the one- and two-electron solutions. The blue- and red-fill areas show the localizations of ME which is free and perturbed by EE, respectively. (inset) $\rho_1 |\tilde{\Phi}_N(\rho_1)|^2 - \rho_1 |R_{1l_1n_1}(\rho_1)|^2$ for $N = 5$ is shown. (bottom) The radial distributions $\rho_{EE}$, $\rho_2 |R_{2l_2n_2}(\rho_2)|^2$ and $\rho_2 |\tilde{\Phi}_N(\rho_2)|^2$, of EE in the disk plane, obtained from the one- and two-electron solutions. The blue- and red-fill areas show the localizations of EE non-perturbed and perturbed by ME, respectively.

## One-electron Diskoid-like States

Here, we describe the diskoid-like states in an one-electron approximation, where we evaluate the electron binding potential classically, rather than by resorting to the semi-classical solutions discussed above. First, we calculate the electrostatic potential of an electron interacting with a perfectly conducting (isolated or grounded, charged or neutral) nanodisk. To this goal, we obtain the analytical solution of a Poisson equation with appropriate boundary conditions. Then, we numerically solve the single-electron Schrödinger equation for the electron moving in this potential. Note that the system is somewhat different than in the previous section, since the charge is freely distributed on the whole disk and the screening is ideal, rather than having a positively charged nucleus and one screening ME moving in 2D.

**Charge distribution on a metallic disk.** We start by finding the equilibrium charge distribution $\sigma(x, y)$ of an isolated metallic disk (centered at the coordinate origin and oriented in the $x - y$ plane) with a radius $a$ and a charge $Q$. The electrostatics of a perfectly conducting uniform thin circular disk was first discussed by Lord Kelvin[31]. He derived the expressions for a surface charge density $\sigma(\rho)$ on an equipotential disk using the formula for the gravitational potential of an elliptical homoeoid.

The Kelvin's approach gives the surface charge density of an ellipsoid,

$$\sigma(x,y,z) = \frac{Q}{4\pi abc} \frac{1}{\sqrt{\frac{x^2}{a^4} + \frac{y^2}{b^4} + \frac{z^2}{c^4}}}.$$

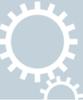

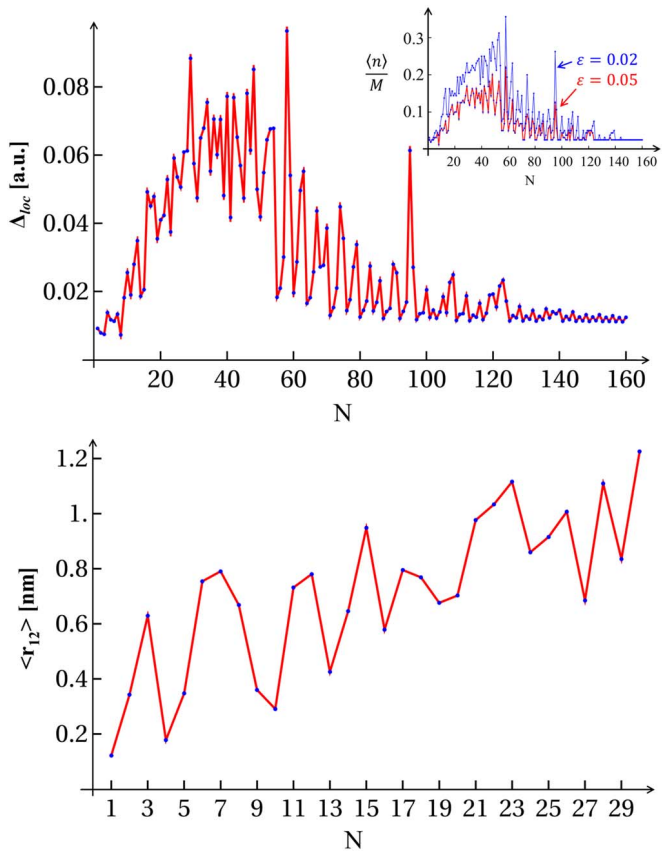

**Figure 5 | (top)** Degree of delocalization of energy eigenfunctions in $N$ states over the basis wavefunctions, $\Delta_{loc}$, where $M = 160$ is total number of basis wavefunctions. **(inset)** The average number of states $\langle n \rangle / M$ whose variational coefficients exceed a threshold of $\epsilon = 0.05, 0.02$. **(bottom)** The average electron-electron distance $\langle r_{12} \rangle$ as a function of $N$.

By projecting this surface charge density on the plane $z = 0$, one can obtain the charge density of an ellipse,

$$\sigma(x,y) = \frac{Q}{4\pi ab} \frac{1}{\sqrt{1 - \frac{x^2}{a^2} - \frac{y^2}{b^2}}},$$

and a disk,

$$\sigma(\rho) = \frac{Q}{4\pi a} \frac{1}{\sqrt{a^2 - \rho^2}}, \quad \rho^2 = x^2 + y^2.$$

In all these cases, the potential is constant over the particular geometrical objects.

The problem of finding the electrostatic potential generated by a charged equipotential disk can be reduced to the equations for the unknown function $f(k)$ (mixed boundary conditions),

$$
\begin{aligned}
V_\sigma(\rho, z) &= \int_0^\infty dk\, f(k) e^{-k|z|} J_0(k\rho), \\
V_\sigma(\rho, 0) &= V_0, \quad 0 \le \rho \le a, \\
\frac{dV_\sigma}{dz}(\rho, 0) &= 0, \quad a < \rho < \infty
\end{aligned}
\tag{3}
$$

where $Q = V_0 \frac{2a}{\pi}$. These equations have the solution, $V_\sigma(\rho, z) = Q \int_0^\infty dk \frac{\sin(ka)}{ka} e^{-k|z|} J_0(k\rho)^{32}$, giving the same expression for the charge density as above.

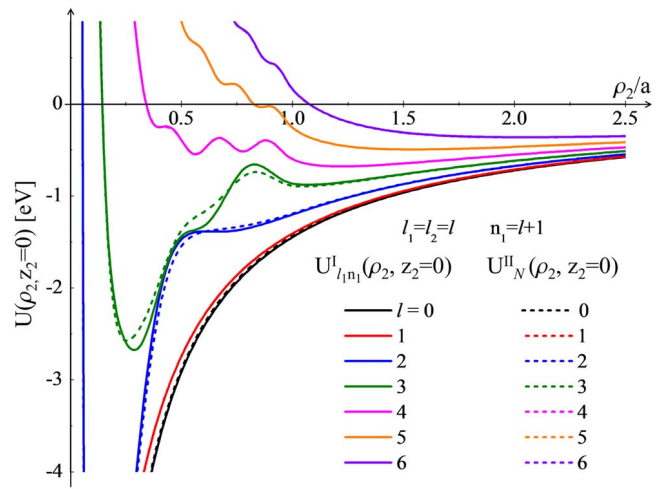

**Figure 6 | Comparison of the mean-field potentials without** $U^I(\rho_2, z_2 = 0)$ **(solid lines) and with** $U^{II}(\rho_2, z_2 = 0)$ **(dotted lines) the ME polarization.** The first is calculated for the states with $l_1 = l_2 = l$ with $n_1 = l + 1$.

**Induced charge distribution and image potential.** Next, we find the potential energy of a point charge above a metallic disk. When a metallic disk is present in a potential created by external charges, $V_{ext}(\rho, \phi, z)$, it develops an induced surface charge density, $\sigma(\rho, \phi)$, and a related potential, $V_\sigma(\rho, \phi, z)$, which makes the total potential $V_{total} = V_{ext} + V_\sigma$ constant on the disk. Depending on the isolated or grounded character of the disk, the total induced charge (integral of $\sigma(\rho, \phi)$ over disk) remains constant (zero) or not. For a neutral isolated disk, $\sigma$ and $V_\sigma$ originate solely in the disk polarization.

Due to the presence of disk edges, finding the potential $V_\sigma(\rho, \phi, z)$ can be a relatively complex task. Since the Lord Kelvin's solution, many sophisticated techniques have been developed. For example, Sommerfeld has shown[33] that such problems can be solved by image potential techniques. In this way, a Green's function for a conducting disk and a point charge with Dirichlet boundary conditions could be found[34,35].

Alternatively, the electrostatic potential of a point charge above a disk[36] could be found from the Copson theorem[37]. The target is to find the electrostatic potential $V_\sigma(\vec{r})$ created by the induced charge density $\sigma(\rho)$ present on the disk, which can be expressed in the external space as,

$$V_\sigma(\vec{r}) = \int_{\substack{\rho' < a \\ z' = 0}} \sigma(\rho') \frac{d^2\vec{r}'}{|\vec{r} - \vec{r}'|}. \tag{4}$$

According to the central lemma, the potential generated by the external charge in the area of the disk can be expressed as,

$$V_\sigma(\rho) = \int_{-\pi/2}^{\pi/2} \lambda(\rho \cos\beta) d\beta, \tag{5}$$

where $\lambda(x)$ is a so called "strip function",

$$\lambda(x) = \int \sigma(x, y) dy. \tag{6}$$

Equations 5 and 6 form a pair of integral transforms equivalent to Eqn. 4. The physical meaning of the lemma is in the connection between the strip function $\lambda(x)$ and the potential created on the disk surface. Once $V_\sigma(\rho)$ is known, one can try to find the strip function which would generate the same (up to the additive constant) but opposite sign potential in the disk area, and cancel $V_\sigma(\rho)$. Thus, the total potential in the disk area would be constant. Along with the unique-ness of the Poisson equation solution, this provides the correct expression for $\lambda(x)$ or $\sigma(\rho)$ and thus, for $V_\sigma(\vec{r})$.

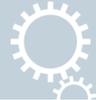

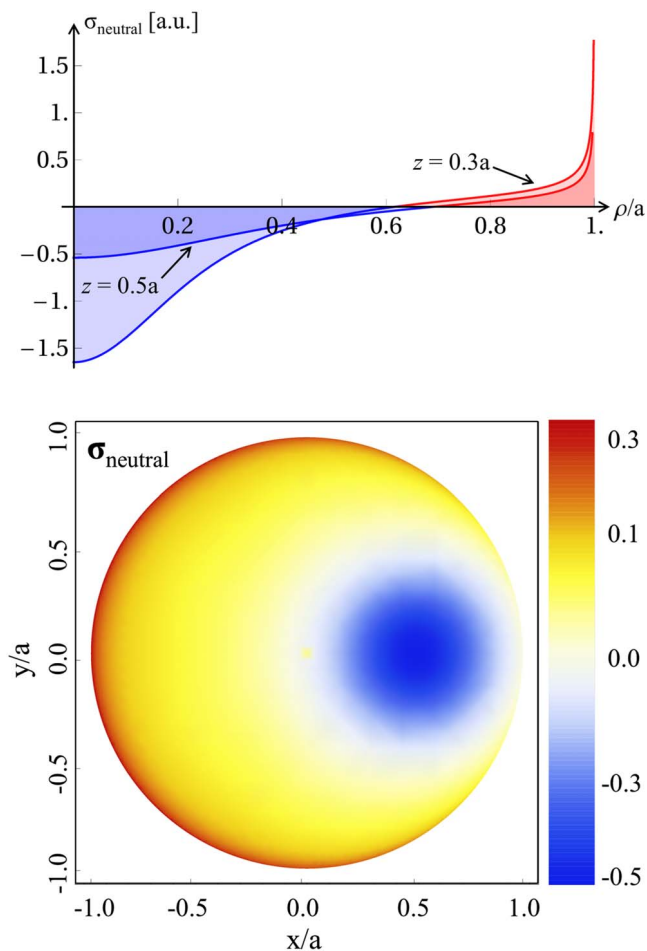

**Figure 7** | (top) Induced image charge densities on a neutral disk for two axial positions $z = 0.3a$, and $z = 0.5a$ of an external point charge. (bottom) Induced charge density on a neutral disk for a point charge positioned above a disk at $\rho = 0.5\ a$, and $z = 0.5\ a$.

In some cases the strip function can be easily guessed. A point charge, $q_{ext}$, placed at $\rho = 0$ and $z = z_0$ creates on the disk the potential $V_\sigma(\rho) = \frac{q_{ext}}{\sqrt{z_0^2 + \rho^2}}$. By inspecting this $V_\sigma$, the strip function can be easily found as $\lambda(x; z_0) = -\left(\frac{q_{ext}}{\pi}\right)\frac{z_0}{z_0^2 + x^2}$. We can use this strip function to compute the interaction potential energy between a point charge placed at the disk axis $\rho = 0$ at distance $z$ above the disk, $V_{\sigma,\rho=0}(z) = -\frac{q_{ext}}{\pi}\left(\frac{a}{z^2+a^2} + \frac{1}{z}\arctan\frac{a}{z}\right) + \frac{2q_{ext}}{a\pi} \times$ $\arctan^2\frac{a}{z}$. For $z \to 0$, we obtain $V_\sigma(z) \sim -\frac{1}{2z}$, while for $z \gg a$, $V_\sigma(z) \sim -\frac{8a^5}{4\pi z^6}$, as expected for a charge above an infinite conducting plane and for a point charge coupled to a quadrupole, respectively. In a similar way, we can also find the induced charge density, $\sigma(\rho; z_0) = -\frac{1}{2\pi\rho}\frac{d}{d\rho}\int_{-\sqrt{a^2-\rho^2}}^{\sqrt{a^2-\rho^2}}\lambda\left(\sqrt{x^2+\rho^2}; z_0\right)dx =$ $-\frac{q_{ext}}{\pi^2}\times\frac{z_0}{(z_0^2+\rho^2)^{3/2}}\left(\arctan\frac{\sqrt{a^2-\rho^2}}{\sqrt{z_0^2+\rho^2}} + \frac{\sqrt{z_0^2+\rho^2}}{\sqrt{a^2-\rho^2}}\right) + \frac{q_{ext}}{a\pi^2}\times$ $\frac{1}{\sqrt{a^2-\rho^2}}\arctan\frac{a}{z_0}$.

In Fig. 7 (top), we show the charge distributions on a neutral disk induced by an external point charge positioned at two different distances above the disk center ($\rho = 0$). The closer the point charge is to the disk plane, the more negative image charge the disk develops beneath it. The induced charge density diverges at the disk edges $\sigma(\rho) \sim 1/\sqrt{a^2-\rho^2}$. Given the disk neutrality, we have $q = 2\pi\int_0^a \rho\ \sigma(\rho)d\rho = 0$.

In order to find a general solution for the potential induced by a point charge in an off-axis position above the disk, we need to solve a Poisson equation with Dirichlet boundary conditions. This leads to a pair of dual integral equations involving Fourier components $f_n(\rho)$ of the external potential, $V_{ext}(\rho,\phi) = \sum_{n=0}^\infty f_n(\rho)\cos n(\phi + \alpha)$, calculated in the disk area. Then, the Fourier components of the potential created by the induced charge on the disk, $V_{n\sigma}(\rho, z)$, can be calculated from the integral equations for the unknown function $\Phi(k)$,

$$V_{n\sigma}(\rho,z) = \int_0^\infty dk\ \Phi(k)e^{-k|z|}J_n(k\rho),$$

$$\int_0^\infty dk\ \Phi(k)\ J_n(k\rho) = f_n(\rho), \quad 0 \le \rho \le a,$$

$$\int_0^\infty dk\ \Phi(k)\ k\ J_n(k\rho) = 0, \qquad a < \rho < \infty. \qquad (7)$$

Instead, we consider the Fourier components $\sigma_n(\rho)$ of the induced charge density, which can be obtained with the help the Abel's integral equation,

$$\int_0^a u\ \sigma_n(u)du \int_0^{2\pi}\frac{\cos(n(\theta+\alpha))}{\sqrt{\rho^2+u^2-2\rho\ u\cos(\theta+\phi)}}d\theta = \\ = f_n(\rho)\cos(n(\phi+\alpha)). \qquad (8)$$

Using the Copson's theorem[37], we can find that

$$\sigma_n(\rho) = -\frac{1}{\pi}\frac{d}{d\rho}\int_\rho^a\frac{t\ S_n(t)}{\sqrt{t^2-\rho^2}}dt,$$

$$S_n(\rho) = \frac{1}{2\pi}\frac{1}{\rho^{2n}}\frac{d}{d\rho}\int_0^\rho\frac{t^{n+1}f_n(t)}{\sqrt{\rho^2-t^2}}dt. \qquad (9)$$

Then, the induced charge density in the disk can be found as $\sigma(\rho,\phi) = \Sigma_n \sigma_n(\rho)\cos(n(\phi+\alpha))$.

In Fig. 7 (bottom), we show the induced surface charge density, $\sigma(\rho,\phi)$, on a neutral disk for a point charge located at $r = 0.5\ a$, $z = 0.5\ a$. We can clearly see the induced negative charge beneath the point charge. The charge distribution also diverges at the disk edge.

The resulting potential created by the induced surface charge density $\sigma(\rho,\phi)$ has the form,

$$V_\sigma(\rho,z) = \int_0^a\int_0^{2\pi}\frac{\sigma(u,\theta)u\ d\theta\ du}{\sqrt{z^2+u^2+\rho^2-2u\rho\cos(\theta)}}. \qquad (10)$$

A grounded disk in an external field (of a point charge) gains beside its polarization also a nonzero induced charge, $q$ (integral of $\sigma$). In order to calculate the induced potential when the disk has a net charge $Q$, we need a add a term $\frac{1}{4\pi a}(Q-q)\Big/\sqrt{a^2-\rho^2}$ to $\sigma(\rho,\phi)$ in Eqn. 10 (Kelvin's formula for a total charge of $Q-q$).

One can easily find that $V_\sigma(\rho, z)$ has a short range asymptotics of $-\frac{a}{\rho^2}\big|_{\rho\to\infty}$ (grounded disk), $-\frac{a^3}{\rho^4}\big|_{\rho\to\infty}$ and $-\frac{a^5}{z^6}\big|_{\rho=0,z\to\infty}$ (neutral disk). Therefore, $V_\sigma(\rho, z)$ cannot support the formation of states localized outside the disk. In order to support such states, the long range asymptotics should have a Coulomb character as well. Therefore, in the remaining calculations, we consider a metal disk (radius $a = 1$ nm) bearing a charge of $Q = |e|$, which can provide the necessary long-range asymptotics.

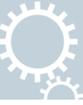

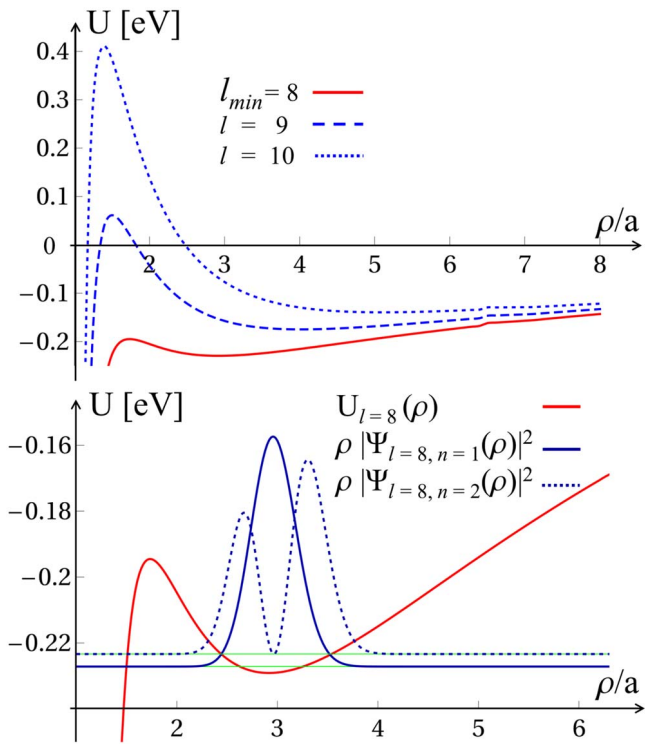

**Figure 8** | (top) Effective potentials $U(\rho, z = 0)$ formed in the disk plane for different angular momenta quantum numbers. (bottom) The lowest two states of an electron ($l = 8$) formed in the effective potential $U$.

**One-electron 1D diskoid-like states.** In order to capture the main features of the diskoids, we consider first 2D motion of an electron in the plane ($z = 0$) of the charged polarizable disk, using the potential obtained in the previous section. The Schrödinger equation for the radial part $R(x) = f(x)/\sqrt{x}$ of the wavefunction is ($x = \rho/a$),

$$-\frac{\partial^2 f}{\partial x^2} + \left(\frac{l^2 - 1/4}{x^2} + \frac{1}{\epsilon}V_{a,z=0}(x)\right)f = \frac{E}{\epsilon}f, \qquad (11)$$

where $\epsilon = \frac{\hbar^2}{2m_e a^2} = 38.1$ meV and $V_{a,z=0}$ is the total electrostatic potential in the disk plane.

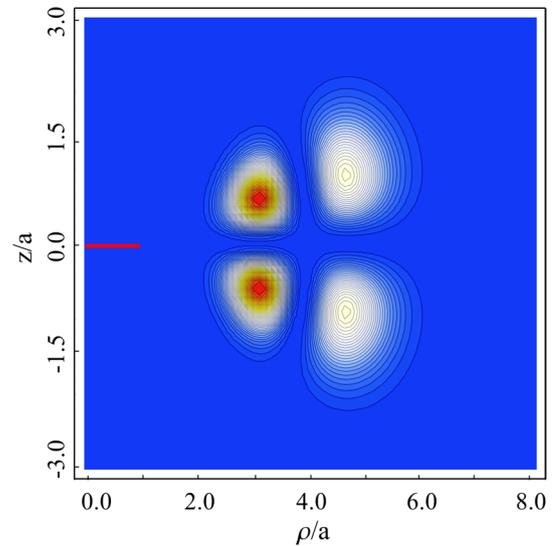

**Figure 10** | **The contour plot for the diskoid state** $|1, 1\rangle$ **with $l = 8$ of a charged disk.** The disk is shown by the red solid line at $z/a = 0$, $\rho/a = 0$–1 middle left.

In Fig. 8 (top), we show the effective potentials, $U$, formed in Eq. 11 by the attractive Coulombic and repulsive centrifugal terms. For $l \geq l_{min} = 8$, the potentials develop local wells, with minima rapidly shifting outwards with increasing $l$. For example, $\rho_{min} = 3.67a$ for $l = 9$. In Fig. 8 (bottom), we also show the ground $n = 1$ and the first excited $n = 2$ states for $l = 8$, with an energy difference of $\Delta\varepsilon \approx 0.004$ eV. In order to disclose the role of polarization in the formation of these extended states, we calculate the states when the disk polarization is removed. In this case, $l_{min} = 5$ and the effective potential has a minimum at $\rho_{min} = 2.4$ $a$, in contrast to $\rho_{min} = 2.9$ $a$ ($l_{min} = 8$) for the conducting disk. These results are similar to those for tubular image states[9–11].

**One-electron 2D diskoid-like states.** Next, we find the full solution of a diskoid-like system described by the 2D Schrödinger equation ($x = \rho/a$, $y = z/a$),

$$-\frac{\partial^2 f}{\partial x^2} - \frac{\partial^2 f}{\partial y^2} + \left(\frac{l^2 - 1/4}{x^2} + \frac{1}{\epsilon}V_a(x,y)\right)f = \frac{E}{\epsilon}f. \qquad (12)$$

The energy factor is $\epsilon = \hbar^2/2m_e a^2$ and $V_a(x, y)$ is the 2D induced electrostatic potential. We solve Eqn. 12 using a finite difference method with the Hamiltonian matrix defined in a 2D square lattice with $(i, j)$ nodes,

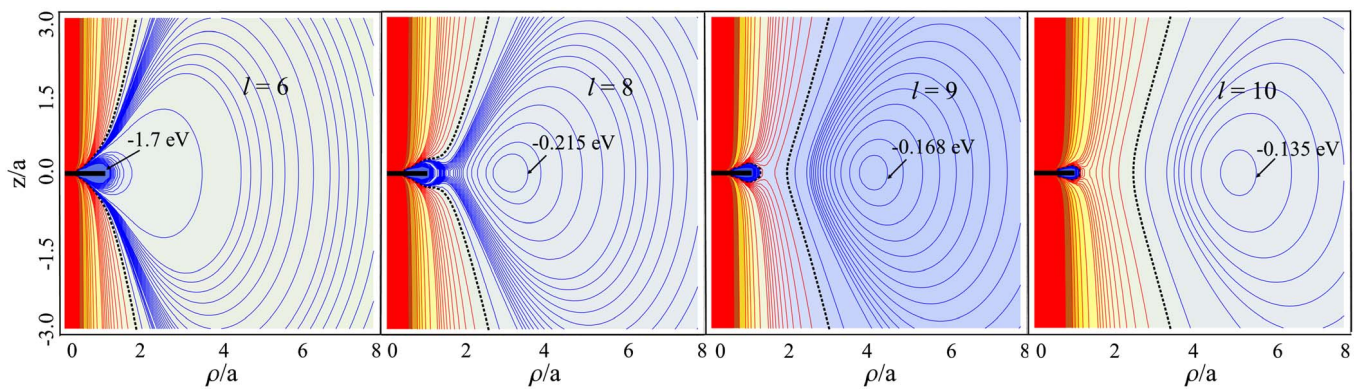

**Figure 9** | **The effective potential of a charged disk** ($a = 1$ nm) for $l = 6, 8, 9$ and 10. The black dashed line marks $U(\rho, z) = 0$. The red area is the repulsion centrifugal wall. The disk is shown by a black solid line.

 **8**

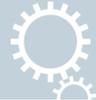

**Table II | Energy spectrum of diskoid-like states (meV)**

| | l | | | |
|---|---|---|---|---|
| n | 8 | 9 | 10 | 11 |
| 1 | −201.75, $\lvert 0,0\rangle$ | −157.81, $\lvert 0,0\rangle$ | −127.75, $\lvert 0,0\rangle$ | −104.70, $\lvert 0,0\rangle$ |
| 2 | −187.50, $\lvert 1,0\rangle$ | −147.67, $\lvert 1,0\rangle$ | −119.71, $\lvert 0,1\rangle$ | −98.03, $\lvert 0,1\rangle$ |
| 3 | −184.53, $\lvert 0,1\rangle$ | −146.69, $\lvert 0,1\rangle$ | −119.51, $\lvert 1,0\rangle$ | −94.84, $\lvert 1,0\rangle$ |
| 4 | −174.73, $\lvert 2,0\rangle$ | −138.28, $\lvert 2,0\rangle$ | −112.11, $\lvert 0,2\rangle$ | −91.15, $\lvert 0,2\rangle$ |
| 5 | −171.84, $\lvert 1,1\rangle$ | −137.46, $\lvert 1,1\rangle$ | −111.45, $\lvert 1,1\rangle$ | −87.49, $\lvert 1,1\rangle$ |
| 6 | −170.77, $\lvert 0,2\rangle$ | −137.06, $\lvert 0,2\rangle$ | −109.86, $\lvert 2,0\rangle$ | −83.23, $\lvert 0,3\rangle$ |
| 11 | −152.71, $\lvert 4,0\rangle$ | −119.44, $\lvert 2,2\rangle$ | −94.72, $\lvert 1,3\rangle$ | −71.34, $\lvert 1,3\rangle$ |
| 12 | −149.96, $\lvert 3,1\rangle$ | −118.92, $\lvert 3,1\rangle$ | −94.34, $\lvert 0,4\rangle$ | −68.30, $\lvert 3,0\rangle$ |
| 13 | −148.75, $\lvert 2,2\rangle$ | −118.52, $\lvert 3,2\rangle$ | −92.65, $\lvert 2,2\rangle$ | −66.00, $\lvert 2,2\rangle$ |

$$\left[\hat{H}R\right]_{\substack{x=i\Delta \\ y=j\Delta}} = (V_{ij}+4t)R_{ij} - t\left(R_{i+1,j} + R_{i-1,j} + R_{i,j+1} + R_{i,j-1}\right). \quad (13)$$

Here, only the nearest neighbors are taken into account and $t=\dfrac{\hbar^2}{2m_e\Delta^2}$ is a "hopping" parameter. The effective potential $U_d(x,y)$ has singularities at $0\le\rho<a$, $z=0$ and $\rho=a$, $z=0$.

In Fig. 9, we show the effective potentials for different $l$, where two distinct types of potential wells are seen to form. For $l\le 7$, the effective potentials develop local minima in the vicinity of the disk edges. An electron orbiting in states localized in these local minima overlap with the disk and easily fall onto its surface. At $l\ge 8$, the competition between a long-range Coulomb attraction ($\sim -\frac{1}{\rho}$) and a centrifugal repulsion ($\sim \frac{1}{\rho^2}$) leads to the formation of an additional minimum (arrow) at $\rho>3a$ and $-3a<z<3a$, where an electron can only cause a negligible polarization of the disk. The formed potential wells are 0.085, 0.368, and 0.915 eV deep (along the path $\rho,z=0$) for the $l=8$, 9 and 10 states, respectively. In Fig. 10, we show one of the states ($l=8$) formed in these wells obtained in the one-electron approximation for a charged and polarizable disk. It has one node in each ($\rho$ and $z$) direction. In Table II, we summarize the energies of these diskoid-like states. Each state is characterized by a pair of quantum numbers $n_\rho$ and $n_z$. The calculations show that the binding energies are approximately twenty to eighty times larger those of tubular image states[11]. At higher $l$ the states change their order.

Note that the potential well formed near the disk edge (Fig. 9) remains there at all $l$'s. This high electric field could potentially be used for the detection and trapping of atoms. At modest charging (1 e-charge on the disk of radius $a=1$ nm corresponds to 2.26 V of external potential) the electric field near the edges could potentially ionize neutral atoms. This effect could be used for the development of compact, cold-atom based interferometers, atom counting and quantum correlation measurements in cold atomic gases[38].

Finally, we would like to briefly discuss the connection between the many-electron and one-electron diskoid-like states, discussed above. In principle, one could find highly excited many-electron (not just two-electron) wavefunctions of metallic nanodisks. In such states, the position of one of the electrons can be fixed and the rest of the many-electron wavefunction can be used to construct a one-electron density matrix. Then, we might find that as the position of the fixed electron is positioned further away from the disk axis, the density matrix produces electron densities similar to that in Fig. 7 (bottom). This does not mean that the hole is dynamically pursuing the electron in a classical sense, but it shows the nature of (static) many-electron eigenstates and their connection to the one-electron solutions. Obviously, in the two-electron solution we can not recover densities with holes perfectly matching the position of the fixed EE in the classical sense, since the single ME can not properly screen this EE.

## Conclusions

We have studied extended diskoid-like electronic states present in highly polarizable molecular or metallic nanosystems, and described their complex correlations and other properties. One-electron (mean-field) solutions have also been discussed and compared to the correlated two-electron solutions. In principle, highly excited DESs in the one-electron limit may be described as "circular" Rydberg-like states with $|m|=n-1$, where $m$ and $n$ are the magnetic and principal quantum numbers, respectively. For large $m$, the orbital of an electron lies in a thin torus centered on the disk axis and reveals quantum position fluctuations around the classical Bohr orbits. These single-electron states have large magnetic moments, smallest Stark effect, and potentially long radiative lifetime[39]. All these features could make DESs useful in quantum entanglement manipulations[40].

However, electron-electron correlations (present in many-electron DESs) do not allow a stable existence of such one-electron states. Therefore, if such one-electron states were somehow prepared, they could decay relatively fast due to correlations; the exchange of angular momentum between electrons (neither $l_1$ nor $l_2$ are good quantum numbers) should decay fast the external electron. Due to the highly correlated nature of the two-electron DESs, they themselves might be prone to coupling and decay, for example, due to light-matter and electron-phonon interactions.

Despite these limitations, DESs are of high fundamental and practical interests, due to many existing nanoscale systems in which they could be observed and applied. These states may be found in the metal nanoparticles[41], metal clusters[42], and graphene flakes[43]. DESs may play a significant role in plasmonic excitations, electron scattering experiments, Wigner molecules, and cold matter science areas[38].

## Acknowledgments

This work was supported by the ACS PRF grant #53062-ND6. A.B. acknowledges the generous support obtained from the Herbert E. Paaren Graduate Fellowship and ITAMP, Harvard-Smythsonian Astrophysical Laboratory. The publication was supported by the Research Open Access Article Publishing (ROAAP) Fund of the University of Illinois at Chicago.

## Author contributions

A.B. carried out all derivations, computations and analysis and wrote the manuscript text. P.K. proposed the project, directed the research, proposed the analyses and edited the manuscript. H.R.S. directed the research and proposed the analysis. All authors discussed the results and commented on the manuscript.

## Additional information

**Competing financial interests:** The authors declare no competing financial interests.

**How to cite this article:** Baskin, A., Sadeghpour, H.R. & Král, P. Correlated Diskoid-like Electronic States. *Sci. Rep.* **4**, 5913; DOI:10.1038/srep05913 (2014).